\documentclass{JINST}
\usepackage{subfig}
\usepackage{graphicx}
\title{Precision Vector Control of a Superconducting  RF Cavity driven by an Injection Locked Magnetron \thanks{ Work supported by Fermi Research Alliance LLC. 
Under DE-AC02-07CH11359 with U.S. DOE.}}

\author{Brian~Chase, Ralph~Pasquinelli,  Ed~Cullerton~ and Philip~Varghese\\
\llap Fermi National Accelerator Laboratory (FNAL),\\
Batavia, IL 60510, USA\\
E-mail: \email{chase@fnal.gov}}

\abstract{The technique presented in this paper
enables the regulation of both radio frequency amplitude and phase in narrow band devices such as a Superconducting RF (SRF) cavity driven by constant power output devices i.e. magnetrons.  The ability to use low cost high efficiency magnetrons for accelerator RF power systems, with tight vector regulation, presents a substantial cost savings in both construction and operating costs  compared to current RF power system technology.  An operating CW system at 2.45 GHz has been experimentally developed.  Vector control of an injection locked magnetron has been extensively tested and characterized with a SRF cavity as the load.  Amplitude dynamic range of 30 dB, amplitude stability of 0.3\% r.m.s, and phase stability of 0.26 degrees r.m.s. has been demonstrated.}


\begin{document}

\section{Introduction}

       RF power sources for accelerators have been based on a variety of technologies including triodes, tetrodes, klystrons, IOTs, and solid-state amplifiers.  The first four are vacuum tube amplifiers; a technology that has been the prime source for powers exceeding hundreds to thousands of watts.  In the past decade, solid-state has become a strong competitor to power amplifiers in the kilowatt(s) power level up to 1 GHz.  All of these technologies have a significant cost that can range from \$5-\$25 per watt of output power.  These same technologies have AC to RF power efficiency potential of close to 60\% in CW saturated operation.
\par
The magnetron is another vacuum tube technology that was developed during World War II for the early radar systems.  Unlike the other devices listed, the magnetron is an oscillator, not an amplifier, but it can be injection locked with a driving signal that is a fraction of the output power\cite{Adler}. The resulting injection gain can be on the order of 15-25 dB.  This gain level is commensurate with IOTs, triodes, and tetrodes.  Klystrons and solid-state can easily achieve gains in excess of 50 dB.  The attractive parameter of magnetrons is the cost per watt of output power.  Magnetrons are the devices used in kitchen microwave ovens, industrial heating systems, and military radar applications.  The cost of a kitchen 1 kW magnetron is under \$10 and simple ready to use ovens are available at under \$100 at this power level.  Industrial 80 kW continuous wave (CW) heating magnetron sources at 915 MHz are commercially available for \$75K. An added benefit to magnetrons is their efficiency.  While alternative technologies approach 60\% efficiency at saturated power output, industrial magnetrons routinely operate at 70\% to 80 \% efficiency level.  This improved efficiency will considerably reduce the operating electricity cost over the life of the accelerator.
\par
For accelerator applications, a high degree of vector control is essential to achieve the required stable accelerating gradient for particle beams.  The magnetron will have an output that is essentially a saturated value for the given voltage and current applied to the device.  Injection locking has been shown to provide a very stable output phase \cite{Wang}, but high dynamic range control of the amplitude cannot be achieved without some additional signal conditioning \cite{Kaz},\cite{Tahir}.
\par
Amplitude control of the cavity field may be achieved by phase modulating (PM) the drive signal to an injection locked magnetron. A predictable amount of power, determined by the phase modulation depth, is displaced from the carrier to sidebands, which are separated from the carrier by the modulation frequency, allowing precise amplitude control of the carrier.  With PM, the ratio of energy in the sidebands to carrier follows coefficients related to Bessel functions.  PM is chosen over other polar modulation techniques such as frequency modulation (FM) because the desired end result is the absolute value of the phase at the cavity as is required for beam acceleration.  These sidebands are designed to be outside the bandwidth of the cavity, so that the sideband power is reflected from the cavity back to a circulator and load, leaving only the desired amplitude regulated carrier and close to carrier information required to drive and regulate the cavity field. The phase of the vector is controlled before the addition of the PM scheme, allowing full quadrature amplitude control of the cavity RF vector.
\par
\begin{figure*}[tb]
\centering
\includegraphics[width=150mm]{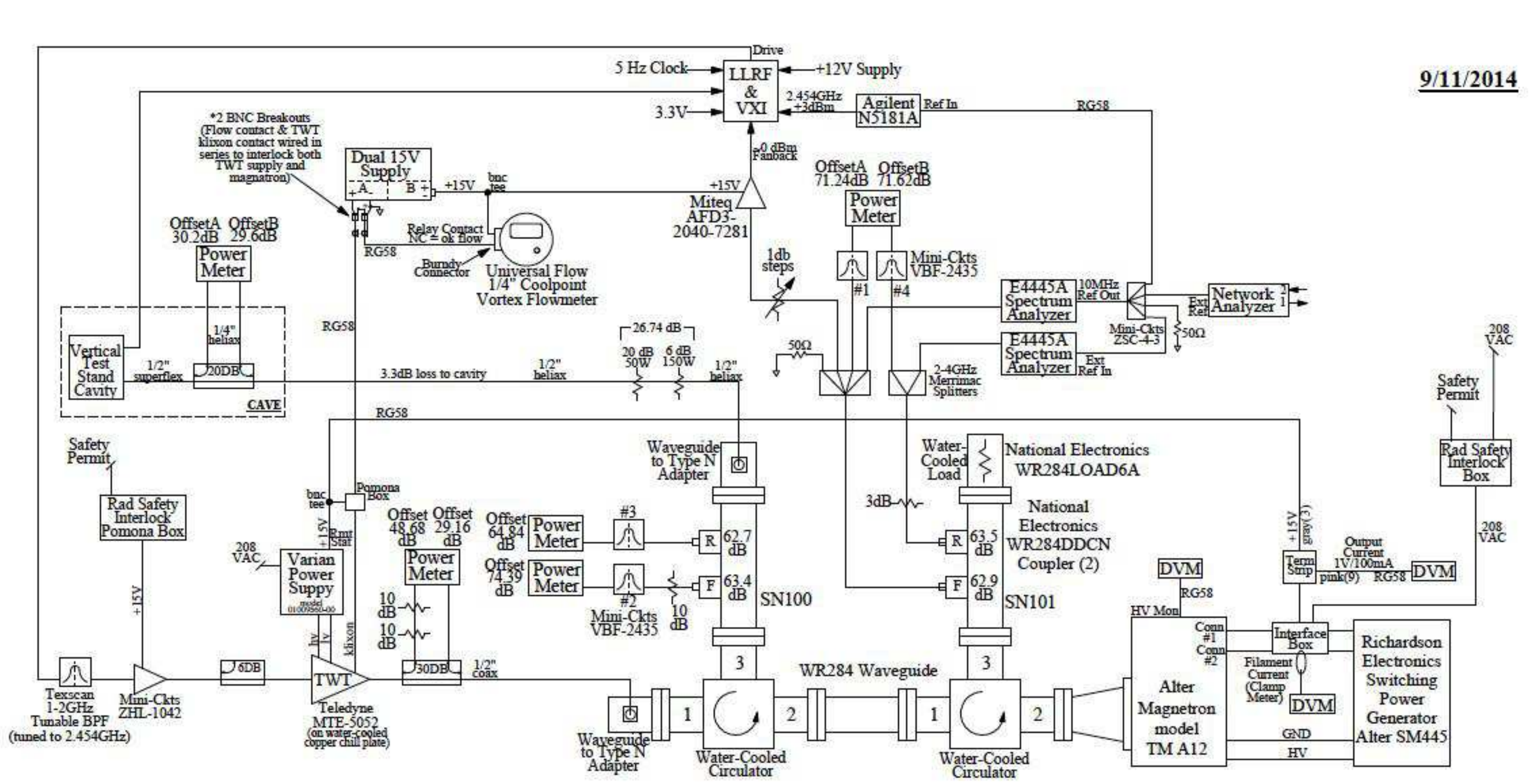}
\caption{RF system configuration}
\label {fig1}
\end{figure*}
\begin{figure*}[tb]
\centering
\includegraphics[width=150mm]{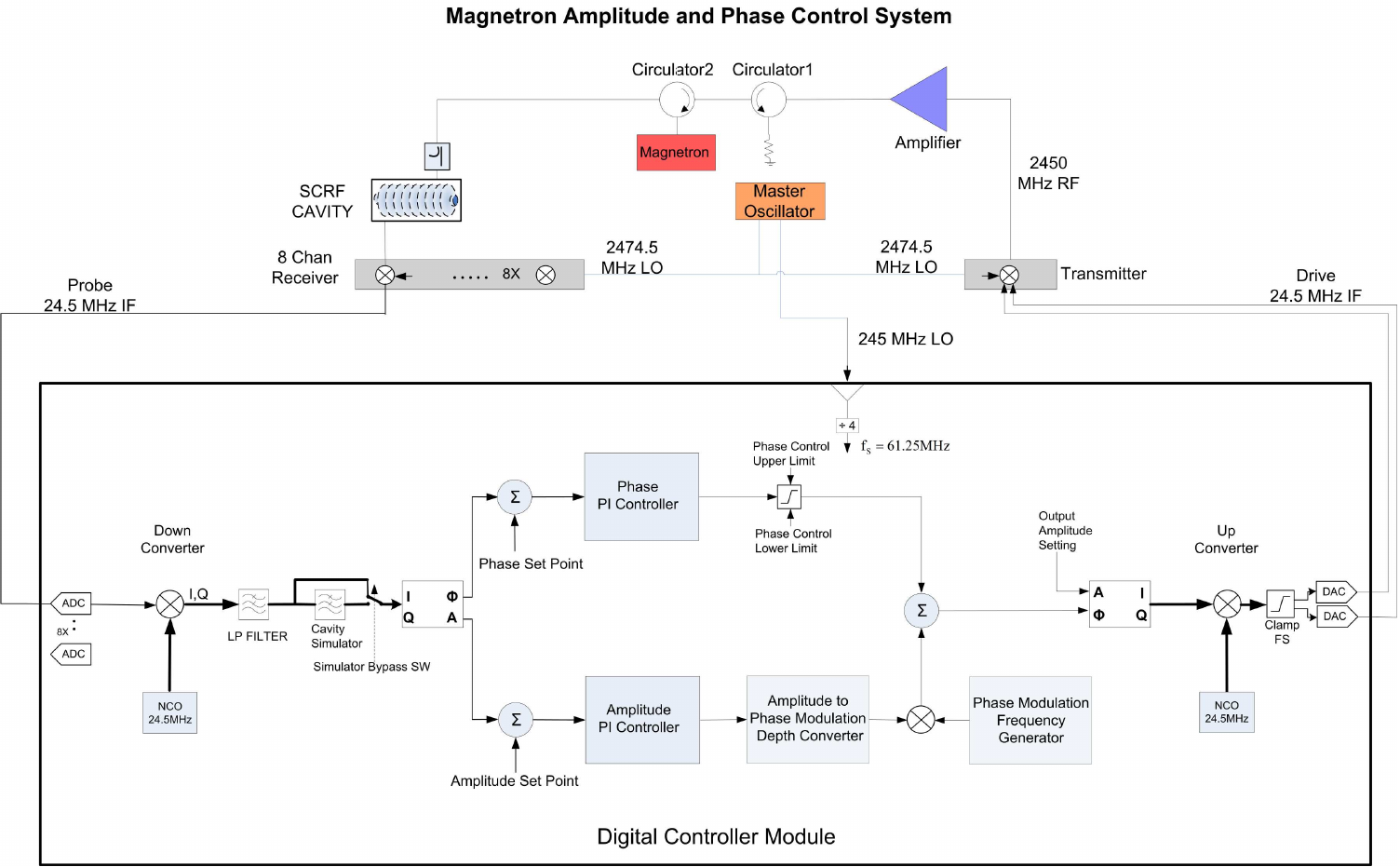}
\caption{LLRF system}
\label {fig2}
\end{figure*}
 
The control scheme, referred to as Low Level RF (LLRF), is implemented using a FPGA baseband digital controller to tightly regulate the amplitude and phase of an RF cavity or other narrow band system with large control bandwidths on the order of 100 kHz. This wideband RF vector control technique may be used in conjunction with lower bandwidth modulation of the magnetron DC power supply (class S) to optimize operational efficiency by keeping the power overhead at a minimal level needed for control. Active control of the magnetic field in the magnetron may be used to keep the magnetron free running frequency centered at the RF frequency.  These combined techniques allow for fast control, low cost and highly efficient operation.  While the generation of sidebands and the reduction of power of the carrier is a well understood and documented modulation technique, it has not been used to control the power delivered to a load such as a SRF cavity from a constant power device such as a magnetron.
\par
  
Using the technique described in this paper, a fully vector controlled power source can be had at a fraction of the cost of alternative methods.  This technique becomes especially attractive for the use with SRF cavities in accelerator applications.  Significant accelerator R\&D worldwide now centers on the use of such high resonant Q cavities. Tens of megavolts per meter of accelerating gradients can be attained with a modest RF drive power.  The cavity acts as a transformer between the RF power amplifier and the accelerating gap seen by the beam.  With loaded Q's ranging from \( 10^6 \) to \( 10^9 \), the cavity bandwidth is very narrow, often in the 10s of hertz.  This narrow bandwidth still allows accurate control of the cavity RF field and energy transfer to the charged particle beam in a very efficient manor, as there is only a tiny amount of energy dissipated by the super-conducting cavity.  Because of the narrow bandwidth of the cavity, the PM sidebands, which may start at a 300 kHz frequency offset, are reflected by the cavity back to the circulator and to an absorptive load.  Due to high levels of reflected power from a SRF cavity under a variety of conditions, circulators are always installed regardless of the RF power source. Hence, all of the reflected power ends up in a well-matched load on the isolated port of the circulator.
\begin{figure*}[t]
\centering
\includegraphics[width=150mm]{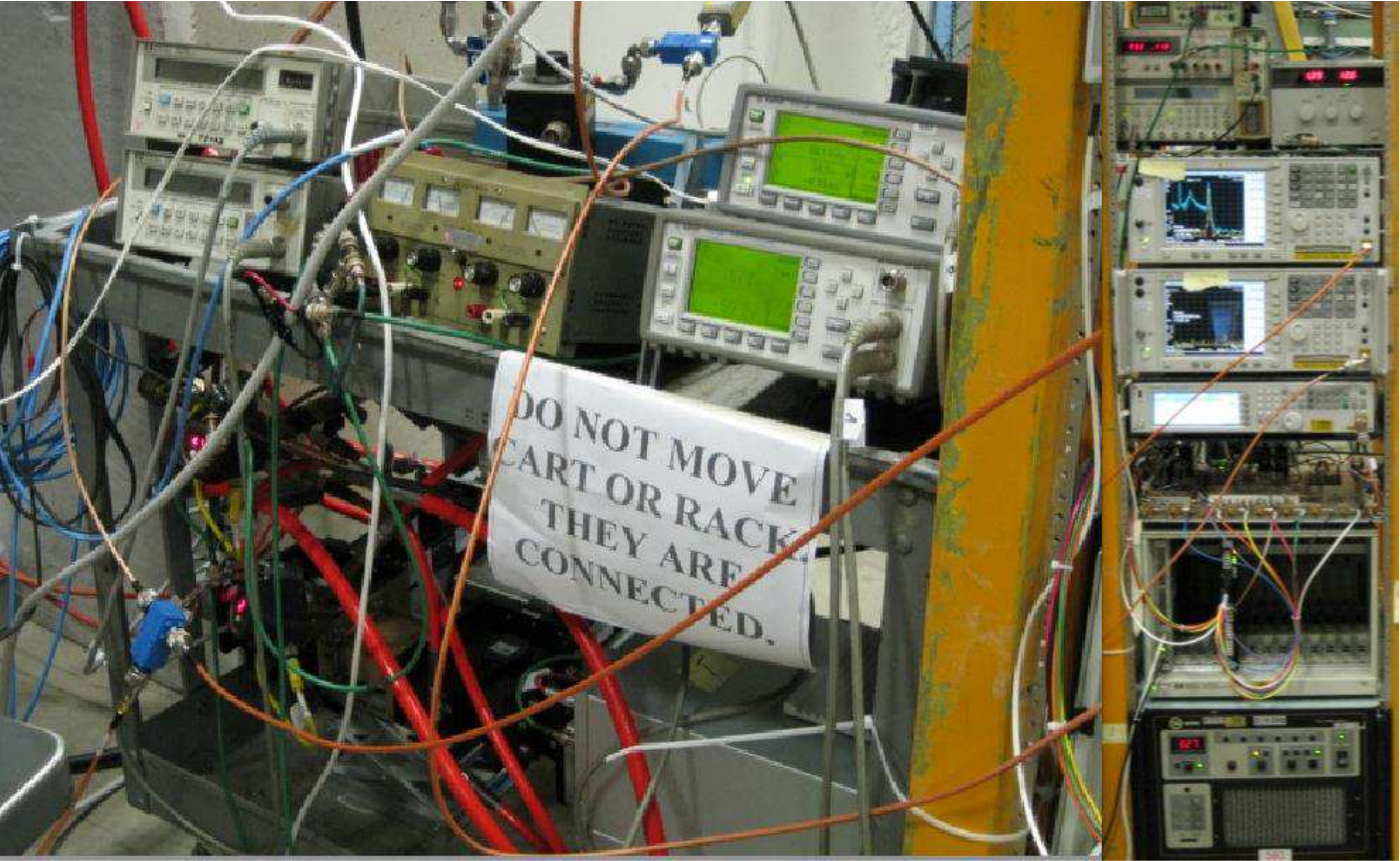}
\caption{Hardware setup for magnetron tests.  Left: cart containing microwave hardware including magnetron, circulators, TWT, and instrumentation.  Right: rack with magnetron power supply, TWT power supply, LLRF, and instrumentation.}
\label {fig3}
\end{figure*}

\par
As previously noted, any polar modulation scheme may be used as long as the RF power device is able to track the phase-frequency waveform and the absolute phase reference is maintained. In this experiment a sine wave modulation waveform is generated in discrete time.  Other waveforms such as a triangle are valid but will require more bandwidth.  Waveforms may be optimized for minimal bandwidth as well.  There are advantages to using this technique even with solid-state power sources.  Current developments with class F amplifiers that are run in hard saturation achieve efficiencies near 90\% but are limited in amplitude bandwidth by the slew rate of the drain power supplies of class S modulation.  Using this wideband polar modulation scheme in conjunction with narrow band class S modulation, a highly efficient wideband system may be achieved.
\par
The experiment presented in this paper utilizes industrial magnetrons at 2.45 GHz,  which is the same frequency used in common kitchen microwave ovens.  The CW saturated output power is 1.2 kilowatts.  This frequency and power level were chosen based on cost and availability of components for this investigation.  Discussions with magnetron vendors indicate that a new design of a magnetron at a specific frequency and power level will have a significant nonrecurring engineering cost for initial development of the magnetron, but that once a proven device is fabricated in quantities of 50 or more pieces, the cost per watt will drop significantly and is expected to be below \$0.50 per watt for the magnetron itself \cite{Private}.  It is estimated that an accelerator based RF system using magnetrons at the 80 kW level is only \$2-\$3 per watt, the added cost over commensurate commercial units is due to the need for a cleaner DC power source and regulation electronics.  This poses an impressive savings over other microwave generators for accelerator applications and is the main purpose for the R\&D.  The cost estimate stated here does not cover the expense of the RF distribution system between the RF source and SRF cavity, which remains the same regardless of the RF power source.  Figure\ref{fig1} and figure\ref{fig2} depict the system block diagrams for this test, figure\ref{fig3} the test hardware.
\section{Efficiency}
\begin{figure}[t]
\centering
\includegraphics[height=60mm]{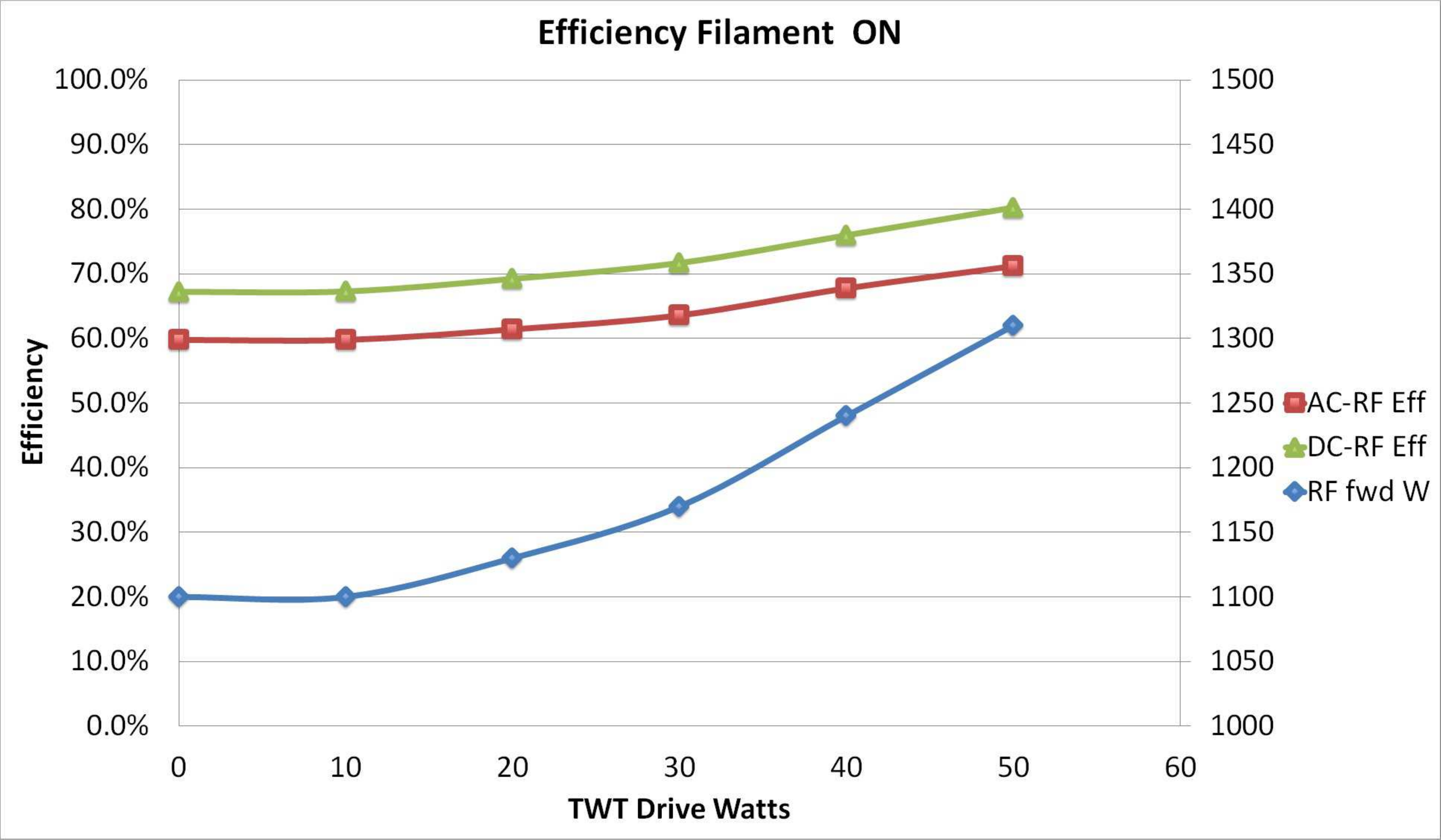}
\caption{Efficiency of injection locked magnetron as a function of injection power with filament on.}
\label {fig4}
\end{figure}

In addition to being a more cost effective solution for accelerator based RF power source, the magnetron is also more electrically efficient than alternative solutions.   The efficiency of the Alter magnetron was measured using a voltage and current probe on the 208 VAC input to the switching supply.  The DC to RF efficiency was also measured by means of a calibrated voltage probes. The power supply was set to 4080 volts and a constant 0.4 amperes.  The output directional coupler was calibrated to an accuracy of 0.1 dB and all other losses were taken into account with an offset enabled on the power meter. In the efficiency test, the magnetron is injection locked with the injection power varying from 1 to 50 watts. If the RF system is operated at 20\% below saturation (minus 1 dB) allowing for some operational overhead, with maximum beam loading, the efficiency will drop to 50\% to 60\% for this device as the reduced carrier power is modulated into rejected sidebands.  The goal for the power systems for SRF is 50\% at this operating level, making the injection locked magnetron a viable solution for accelerator systems.  Figure\ref{fig4} shows the efficiency measured for this magnetron.

\section{SRF Test Results}
After a significant effort of bench testing the injection locked magnetron system, a 2.45 GHz SRF cavity was borrowed from Thomas Jefferson National Lab.  Jefferson Lab constructed this cavity for magnetron tests described in reference \cite{Adler}.  For our test, the coupling on the cavity was set for a Q of \( 10^7 \), a value commensurate with parameters for the next generation SRF linear accelerator at Fermilab.
\par  
The apparatus used is model SM445G 1.2 kW CW magnetron head and power supply from Alter Power Systems, a subsidiary of Richardson Electronics Inc.  Circulators, loads, and directional couplers were purchased from Richardson Electronics Inc.   The signal source is an Agilent N5181A generator with a frequency modulation (FM) input. The power supply is a 21 kHz Alter switching supply.  Unless otherwise noted, all tests were performed with a power supply voltage of 4455 volts, 0.189 amperes.  A TWT was utilized as the injection RF power source due to its availability.  (For future applications it is likely a narrow band solid-state amplifier could be used for the injection RF power source.)  The VXI digital FPGA LLRF controller is a proprietary Fermilab design.
\begin{figure}[t]
\centering
\includegraphics[width=110mm]{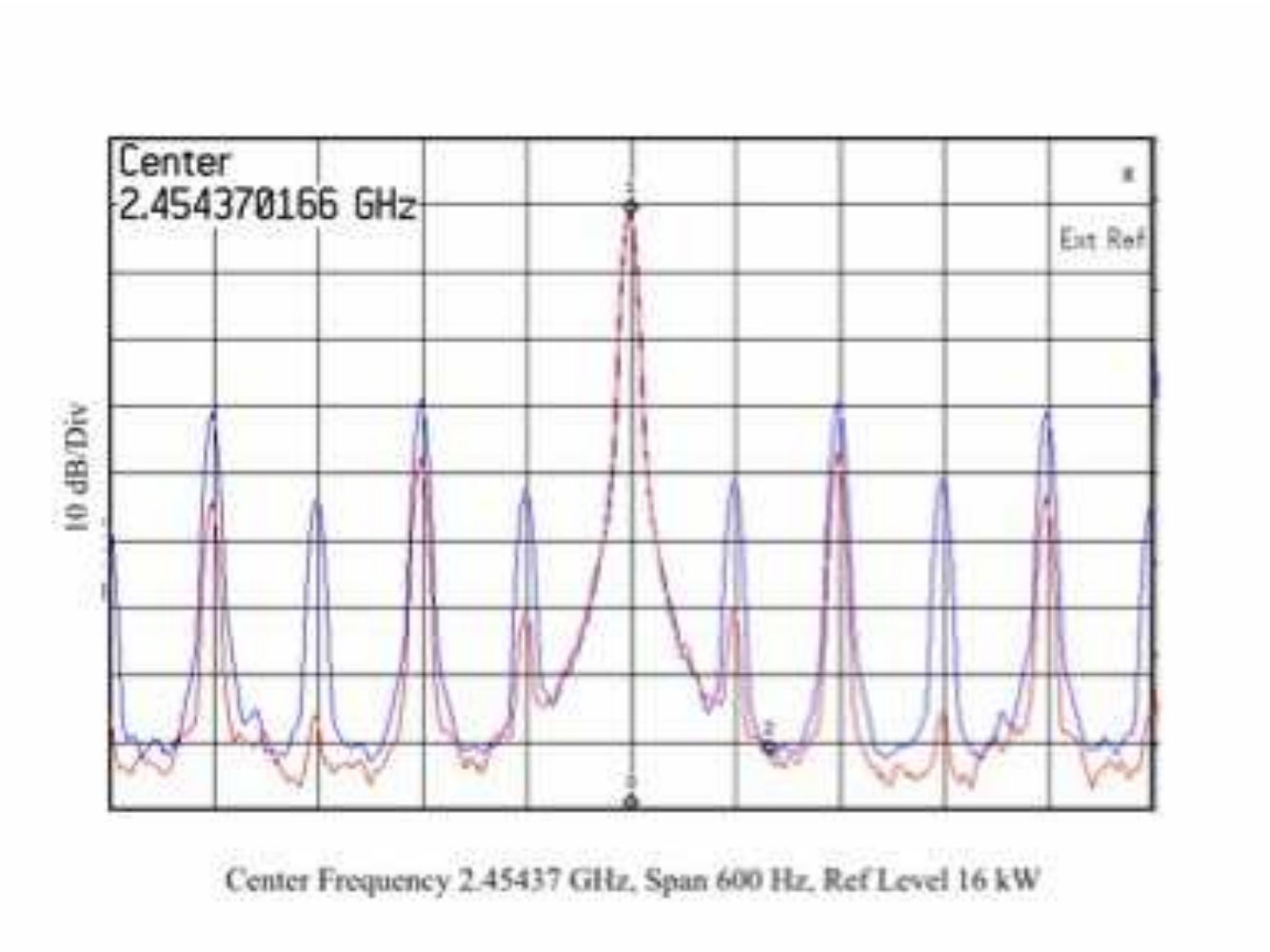}
\caption{Narrow band spectrum of injection locked magnetron. Blue filament on, red filament off. Note suppression of 60 Hz sidebands caused by filament power supply.}
\label {fig5}
\end{figure}

\par
The magnetron is highly susceptible to all harmonics of the switching supply.  For use with a superconducting cavity, this sideband spectrum is typically outside the cavity bandwidth, but the LLRF feedback system does a very good job of suppressing these sidebands.  Potentially more damaging sidebands are the harmonics of the 60 Hz AC filament supply at 3.3 V and 0.11 amperes.  In CW operation, it is possible to run the magnetron with the filament supply turned off after a brief (seconds) warm up.   Figure\ref{fig5} shows the output spectrum with filament on and off with injection locking.  While these tests are mainly concerned with CW operation, the method employed here for full vector control is also applicable to pulsed magnetron operation.  In that case, filament current must always be applied to the magnetron.  A DC filament supply will alleviate any line related harmonics from appearing on the output spectrum. The coupling adjustment to the cavity lowered the resonant frequency to 2.44875 GHz at 4.2 kelvin.  This value is approximately 4 MHz below the natural frequency of the magnetron operating at full power of 1.2 kWatts. 

\par
The Alta power supply is current regulated. Adjustment of the current down to 0.189 amperes generates 500 watts in the magnetron, but more importantly brings the free running frequency of the magnetron closer to the 2.448 GHz necessary for injection locking to the  JLAB cavity frequency.  DC to RF efficiency is 60\% under these operating conditions.  The TWT drive signal is set to 1.5 watts for an injection gain of 25 dB.
\begin{figure}[t]
\centering
\includegraphics[width=90mm]{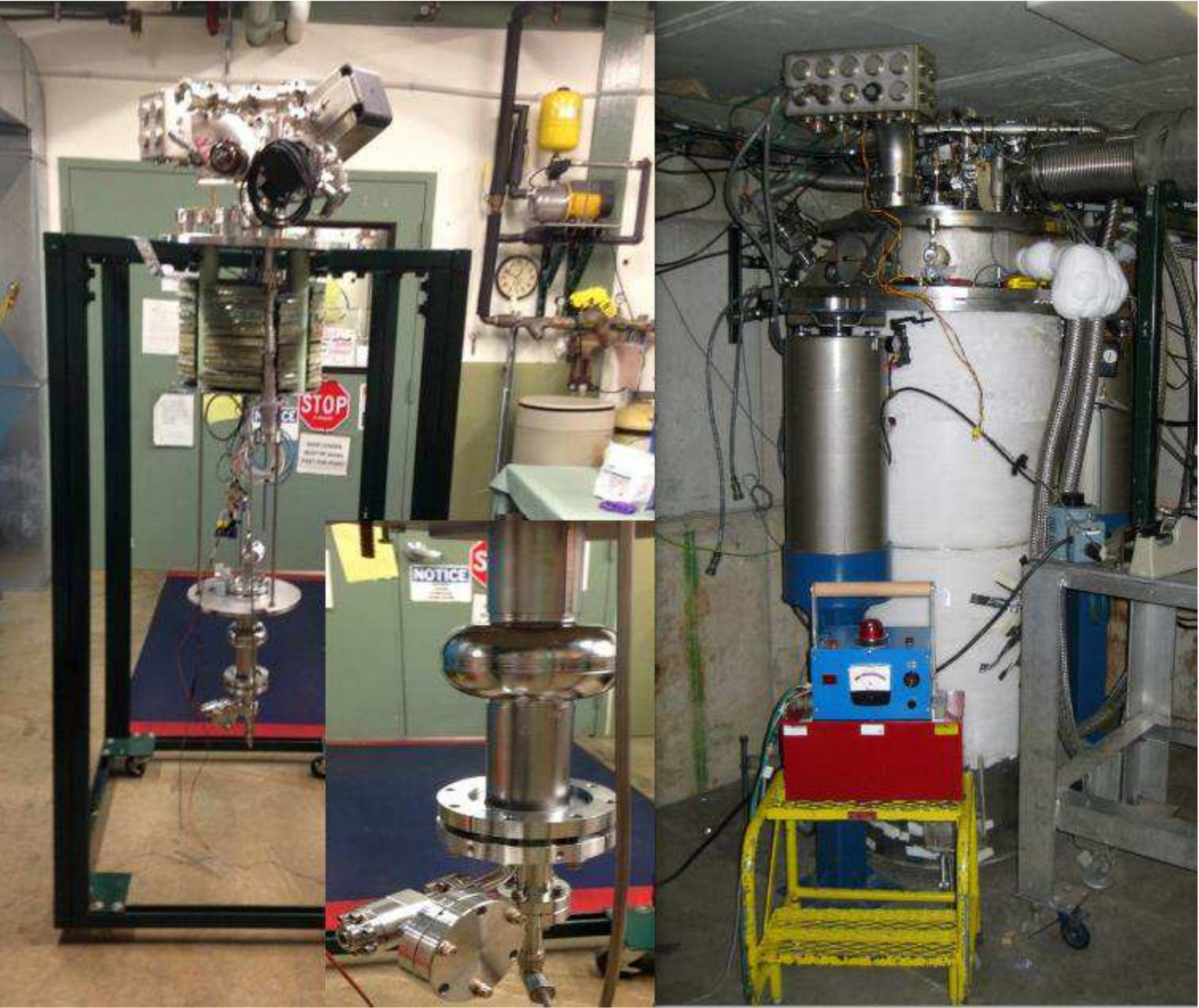}
\caption{Left: Cavity (inset) installed on Dewar insert.  Right: Vertical test stand at A0.}
\label {fig6}
\end{figure}

\par
Since the full power of the magnetron is not needed to achieve the desired RF field in the cavity without particle beam loading, the cavity drive signal was tapped from the input circulator's reflected power port 3.  A 150 watt 6 dB pad and 3.3 dB coax cable loss to the cavity resulted in a cavity drive signal of 1.7 watts. The modulation frequency for the PM is 300 kHz.  All tests of driving the SRF cavity with the magnetron take place under these conditions.
\par
\begin{figure}[b]
\centering
\includegraphics[width=90mm]{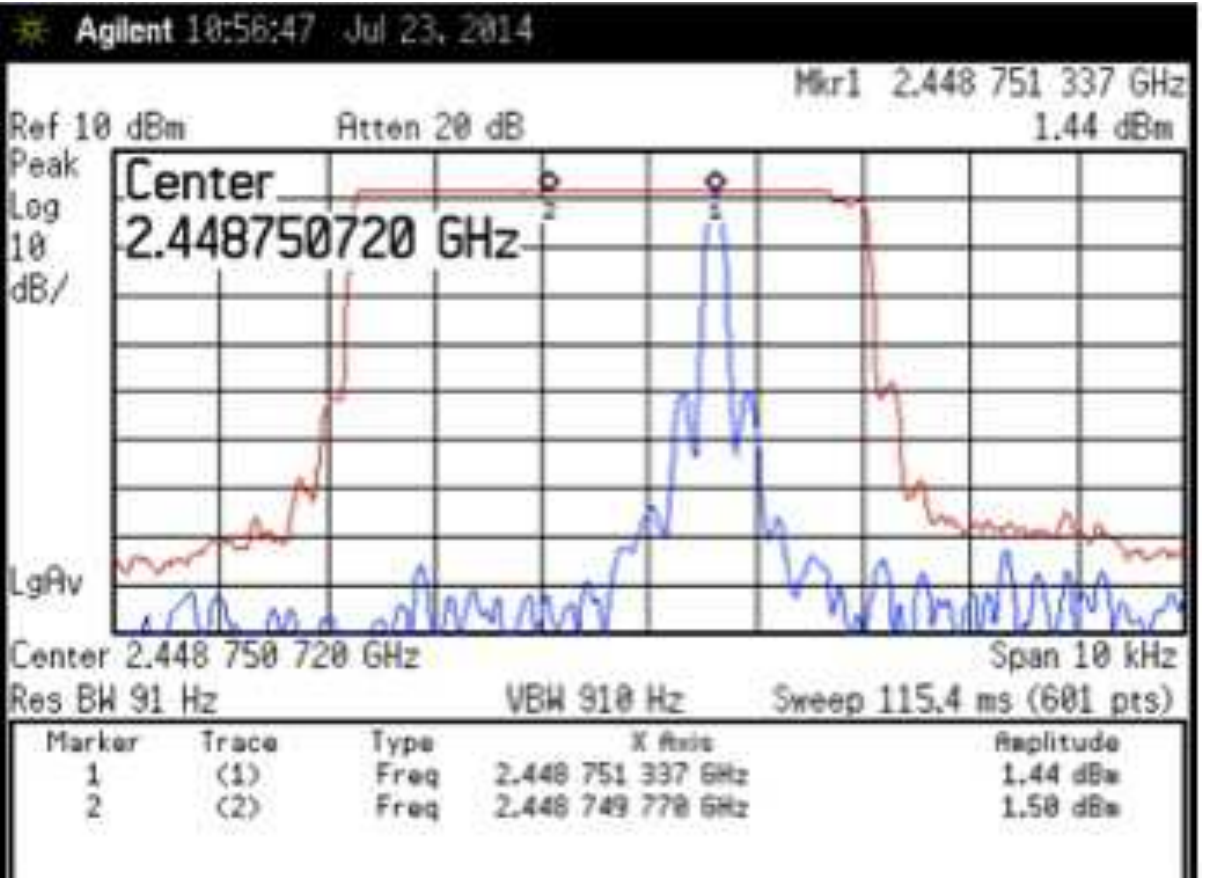}
\caption{Variation of resonant frequency after the Dewar is filled to 90 cm.   Blue is the live trace; Red is in max hold showing the extent of the variation of almost 5 KHz before settling down to under 200 Hz.}
\label {fig7}
\end{figure}
\begin{figure}[t]
\centering
\includegraphics[width=90mm]{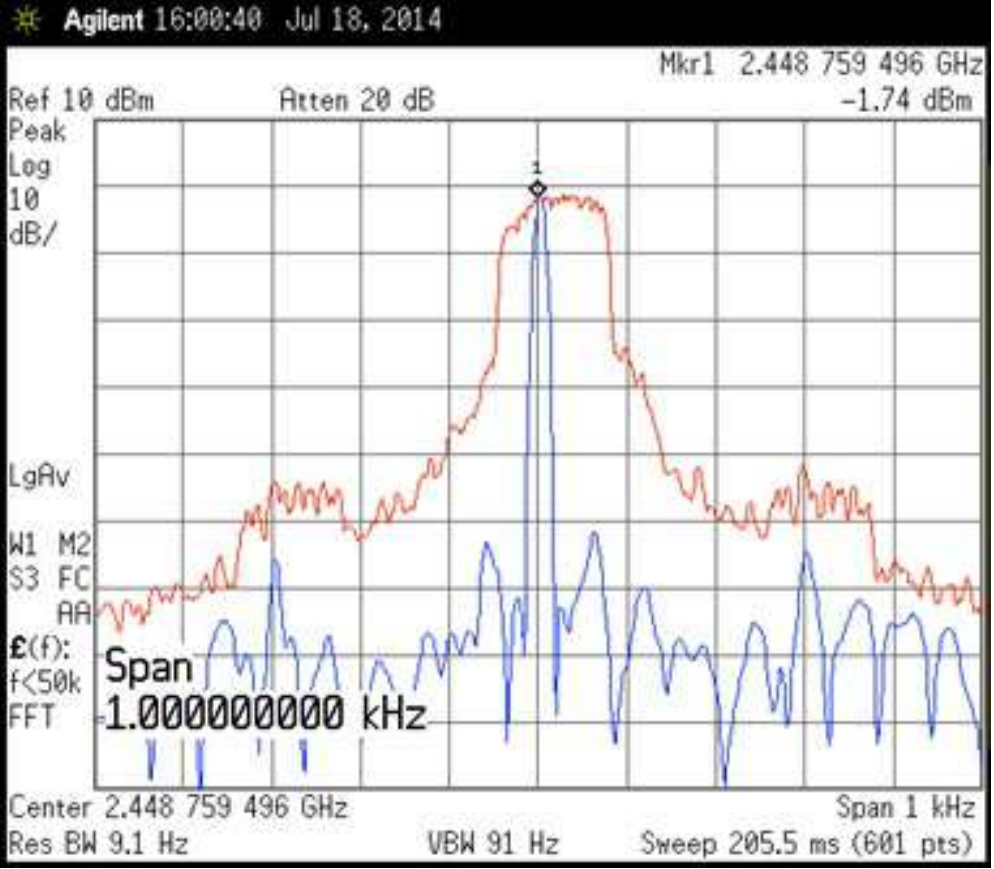}
\caption{Once cavity is stabilized, the FM loop was turned off and rough measurement of the cavity Q was possible by adjusting the frequency of the signal generator.  Blue is the active trace; Red is max hold that traces the bandwidth of the cavity.  The 3 dB bandwidth is on order of 125 Hz.  With the center frequency of 2.44879 GHz this renders a Q of close to 2x\( 10^7 \).}
\label {fig8}
\end{figure}

 Figure\ref{fig6} depicts the Dewar containing the cavity in the vertical test stand at the A0 facility at Fermilab.  With 90 cm of helium in the vessel, the Dewar is at maximum capacity of liquid.  At 4.2 kelvin, boil off of the liquid is at a rate that allows for a minimum of 36 hours before the cavity is no longer immersed.  Upon filling the Dewar, the turbulence in the bath is quite noticeable by monitoring the resonant frequency of the cavity, figure\ref{fig7}.  The pressure fluctuations with this initial condition cause significant variation in the cavity resonant frequency.  A FM loop around the signal generator and cavity fan back was implemented to track the resonance frequency of the cavity with the RF system.  After an hour or two, with fill valves closed, the pressure settles to a value of 775 torr and motion of the resonant frequency is limited to a few hundred Hz.   Figure\ref{fig8} depicts the approximate 3 dB bandwidth of the cavity and resulting Q.

\par
The A0 test stand has the ability to lower the bath temperature to 2 kelvin.  This is achieved with three Roots blower vacuum pumps.  After two hours of pumping, a pressure of 23 torr is achieved and regulated, resulting in 2-kelvin operation.  At this pressure, the frequency of the cavity is 2.448905592 GHz, an increase of 155 kHz.  The pressure sensitivity of the cavity is 207 Hz per torr.  Unfortunately, the vacuum pumps are located only ten feet from the Dewar and the mechanical motion coupling to the Dewar caused excessive microphonic disturbances that did not disappear after two hours of investigation.  The consumption of liquid He under these conditions is considerable.  The A0 test stand is fed from 500 liter Dewars that must be manually replaced as they empty.  It was decided to warm back up to 4.2 kelvin to prolong the testing period for the balance of the experiment.
\section{Low Level RF(LLRF) Controller}
The block diagrams of the RF system are shown in figure\ref{fig1} and figure\ref{fig2}.  The RF signal path may be traced through several sections, first a super-heterodyne 8-channel microwave receiver that down-converts the 2.45 GHz cavity probe signal to a 24.5 MHz intermediate frequency (IF), followed by an analog to digital converter and digital receiver that converts the IF to a baseband analytic signal within a Field Programmable Gate Array (FPGA). The complex In-phase and Quadrature signals (I/Q) may be sent through or bypass around the cavity simulator  before being  converted to amplitude and phase by a CORDIC block.  These amplitude and phase signals are then input to the respective error summing junctions of two proportional-integral (PI) feedback controllers.  The amplitude controller output drives a PM to AM linearizing block, creating a phase modulation depth control signal that is multiplied with a sine wave of a programed frequency.  This now amplitude-controlled sine wave is summed with phase shift request of the more traditional phase control providing the phase modulation input to the second CORDIC block.  The amplitude input of the CORDIC is a settable parameter that is held constant during operation. This is the key difference from a traditional LLRF system as instead of simple amplitude modulation of the RF, the amplitude PI controller controls the phase modulation depth of the signal of a sinusoidal phase modulator of fixed frequency. Modulation frequencies from 100 kHz to 500 kHz have been tested.  A lookup table linearizes the relationship between amplitude request and modulation depth request. The in-phase and quadrature term outputs of the CORDIC are digitally up-converted back to the IF frequency before being converted back to analog and then up-converted  from IF back to RF. The output drive is then a phase modulated by the sum of the phase controller and the sinusoidal phase modulator.  This LLRF drive signal is amplified and then injected into the magnetron, which sympathetically frequency and phase locks to the drive.  The magnetron output signal is directed by the circulator to the cavity and contains all the PM generated sidebands generated by the LLRF system.  The center frequency signal now contains only the intended amplitude signal as requested by the AM PI controller and the phase information requested by the PM PI controller.  The PM sidebands are spaced out in multiples of the phase modulator frequency and are rejected by the narrow band cavity back to the circulator and are terminated by the load.  The cavity probe signal is returned to the LLRF system and is used as the feedback path signal.

The look up table (LUT) for phase modulation depth required for a particular amplitude request is calculated as follows. A phase modulated carrier signal can be written as a sum of an amplitude modulated carrier and modulation sidebands given by
\[
A\cos(\omega_C t + \beta \sin \omega_M t) = A J_0 (\beta) \cos\omega_C t +
\]
\[
 \sum_ {k=1} ^ {\infty } J_{2k} (\beta)[ \sin(\omega_C + 2k\omega_M )t + \sin(\omega_C - 2k\omega_M )t] +
\]
\begin{equation}   
\sum_ {k=0} ^ {\infty }  J_{2k+1} (\beta)[ \cos(\omega_C + (2k+1)\omega_M )t - \cos(\omega_C - (2k+1)\omega_M )t]
\end{equation}
where \( \beta \) is the modulation depth in radians, \(\omega_C \) is the carrier frequency and \(\omega_M \) the modulation frequency. The amplitude modulation of the carrier is given by the Bessel function \( J_0( \beta ) \) which has values of  \(J_0 \)(0) = 1 and \(J_0 \)(2.405) = 0. \(J_0 (\beta) \) can be computed from the first few terms of its series representation
\begin{equation}   
J_0(\beta) = 1-  {{\beta^2} \over {2^2}} + {{\beta^4} \over {2^2.4^2}}- {{\beta^6} \over {2^2 . 4^2 . 6^2}} + \dots
\end{equation}
The amplitude modulation to phase modulation LUT can be generated by numerically solving for \( \beta \) in the equation
\begin{equation}   
J_0(\beta) - \alpha = 0
\end{equation}
for equal interval values of \( \alpha \) from 0 to 1 representing the complete range of amplitude modulation. Equal interval tabulation of the amplitude values allows for an efficient interpolation algorithm that can be completed in 2 cycles.
\section{Closed loop noise performance}
\begin{figure}[b]
\centering
\includegraphics[width=90mm]{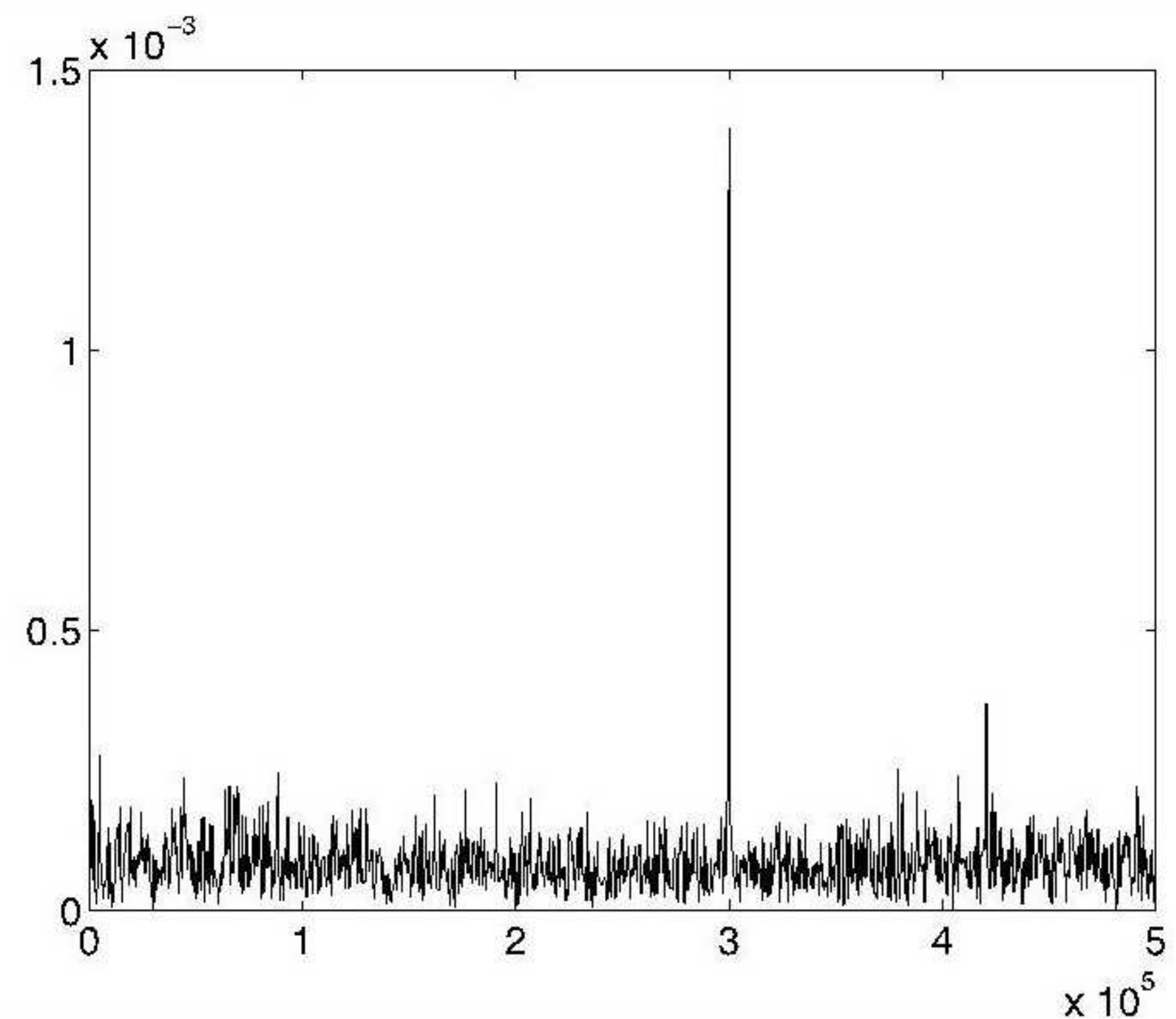}
\caption{FFT of cavity probe error signal with amplitude loop closed.  PM modulation at 300 kHz is seen with 0.14\% residual error.}
\label {fig9}
\end{figure}

The data acquisition by the LLRF system for these studies provides data at 1 MSPS for 10 groups of 2048 samples.  An FFT of the amplitude error signal is shown in figure\ref{fig9}.  What is visible is excellent rejection of line related noise and 21 kHz harmonics of the switching power supply noise.  The large line at 300 kHz is from the phase modulation process and also the 600 kHz line aliased back to 420 kHz.  The rms error for the full signal is 0.30\%, however, if we limit the noise integration to 36 kHz, which is the closed loop bandwidth for the cavity with a 122 Hz half bandwidth and a proportional gain of 300, the rms error drops to 0.15\% including the 300 kHz line.  Many accelerator designs such as PIP-II at Fermilab or LCLS-II at SLAC have cavities with half bandwidths of 15 Hz.  These cavities would reject the 300 kHz modulation by an additional 18 dB.  Also, the phase modulation is a digitally generated signal that allows cavities to be programmed such that the modulation errors are effectively subtracted by the integration process of beam through the cavities.  Given these factors, a 0.01\% rms error appears an achievable goal for narrow band cavities.

\begin{figure*}[t]
\centering
\includegraphics[width=90mm]{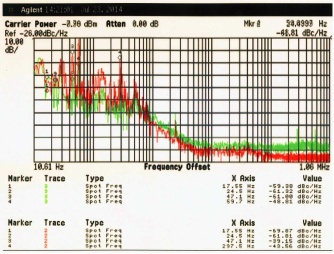}
\caption{Phase noise injection locked magnetron FM loop off. Red: phase and amplitude loops off, Green: phase and amplitude loops on.  Of particular note is suppression of 47, 60, 300 Hz lines.}
\label {fig10}
\end{figure*}
\begin{figure*}[b]
\centering
\includegraphics[width=90mm]{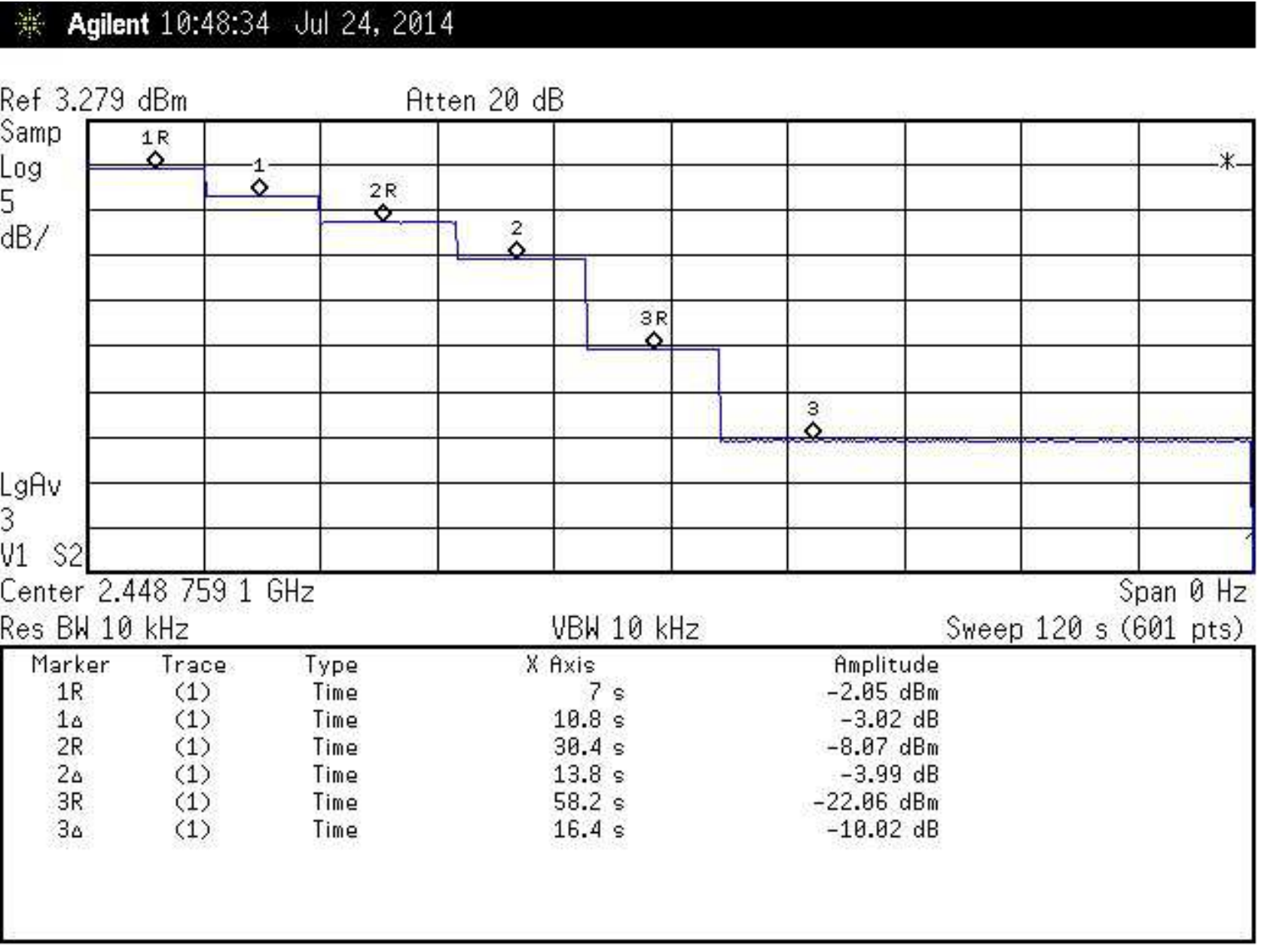}
\caption{Dynamic range with PM modulation.  Cavity probe signal with spectrum analyzer in zero span mode for steps of 0, 3, 6, 10, 20, and 30 dB.}
\label {fig11}
\end{figure*}

\par
The requirement of the FM loop to track the cavity, as a result of the cavity microphonics, in this test conflicts with the operation of the phase loop.  This is solely a function of this test facility as an undressed cavity at 4.2 kelvin in a vertical Dewar is very susceptible to microphonic disturbances that are not evident in a cryomodule. Once the system pressure settled with the FM loop disabled, the phase loop could be enabled over a brief period before the cavity wandered off the drive frequency.   Figure\ref{fig10} shows the phase noise improvement of the system with loops on and off.  Even with a low gain in the phase loop the r.m.s. error is 0.26 degrees.
\par
\begin{figure*}[t]
\centering
\includegraphics[width=3.1in]{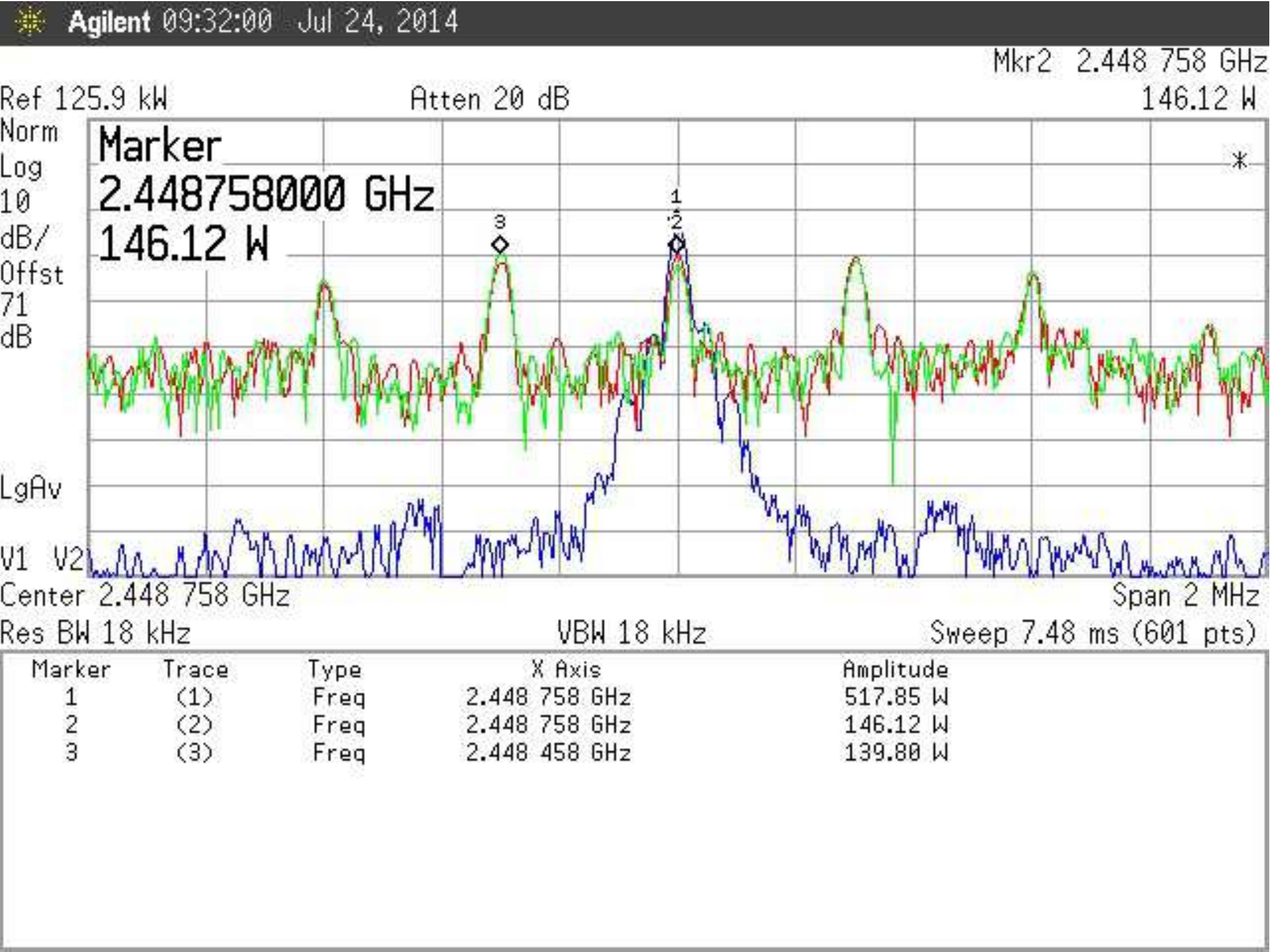}
\caption{Magnetron drive signal, FM loop on all three traces.  Blue all feedback loops off, Red just amplitude loop, Green both amplitude and phase loops on.}
\label {fig12}
\end{figure*}
\begin{figure}[h]
\centering
\subfloat[0 dB (blue), 3 dB (red), 6 dB (green)]{\label{fig12a}\includegraphics[width=0.47\textwidth]{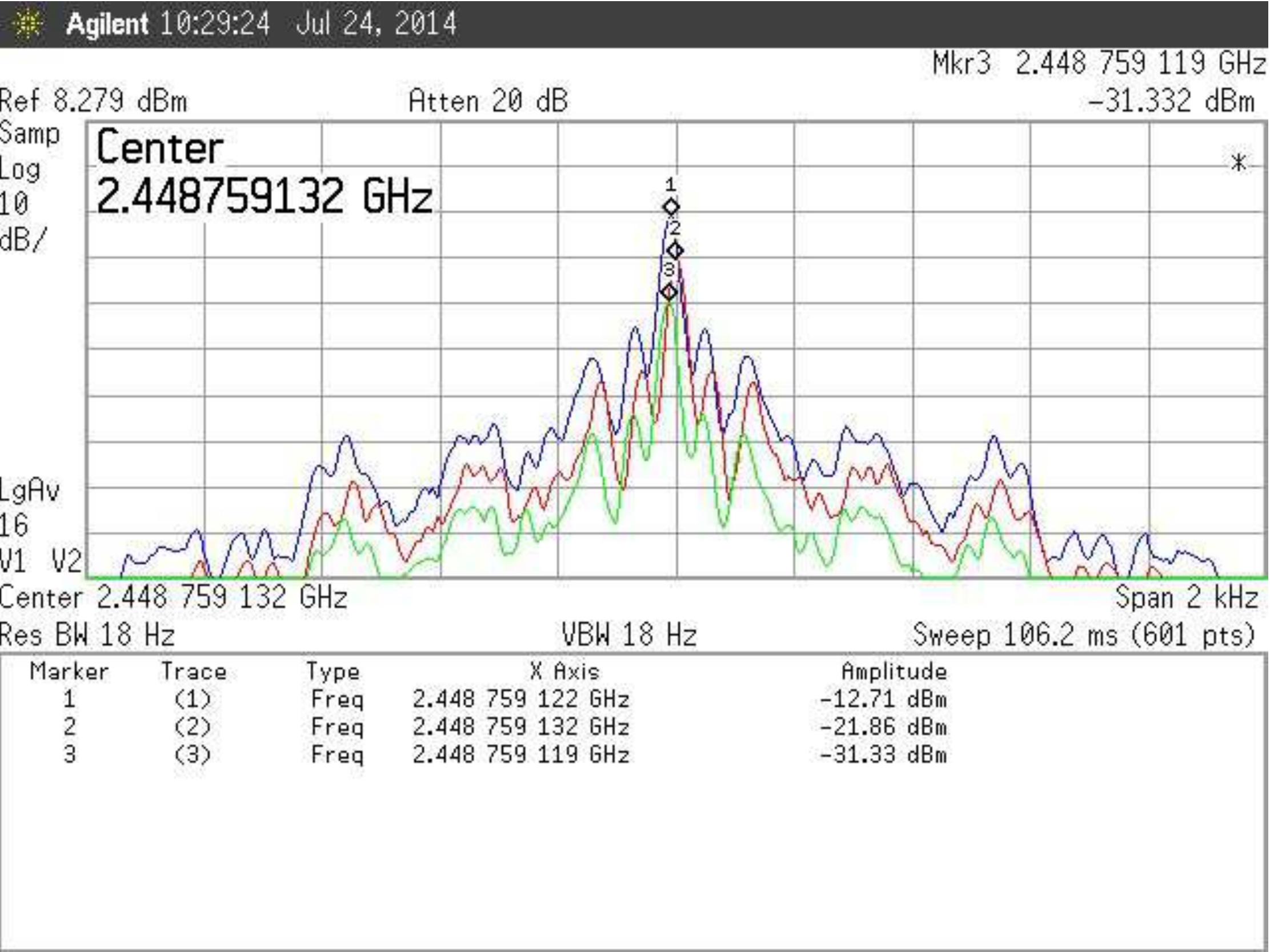}}
\quad
\subfloat[10 dB (blue), 20 dB (red), 30 dB (green)]{\label{fig12b}\includegraphics[width=0.47\textwidth]{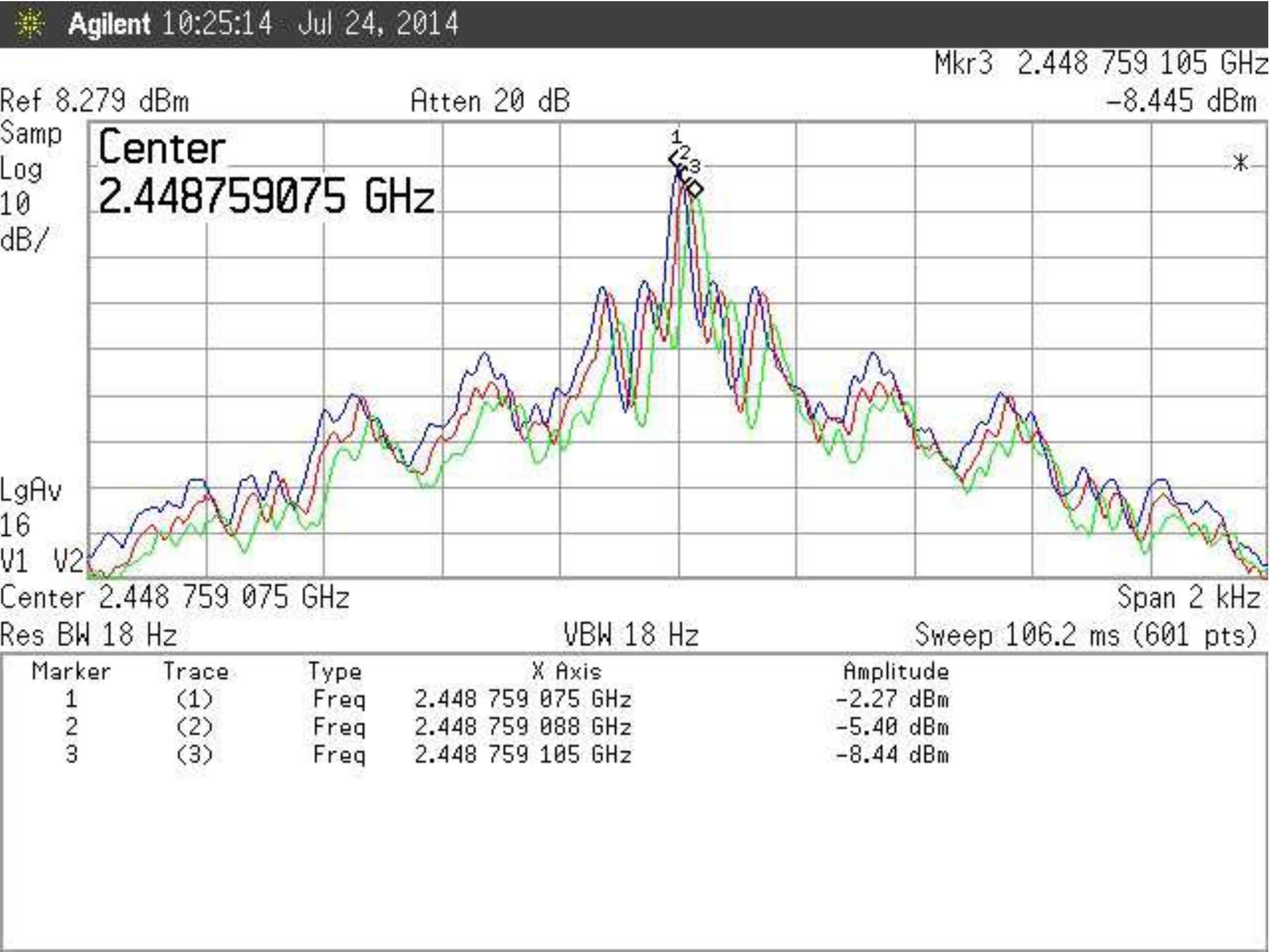}}
\caption{Dynamic range: Steps in the amplitude feed back loop}
\label{fig13}
\end{figure}

Closed loop dynamic range was shown to exceed 30 dB by adjustment of the amplitude loop set point, figure\ref{fig11}.  Figure\ref{fig12} is the spectrum of the drive signal for loops off and on.  The noise floor is significantly higher with loops closed, but the increased power spectral density is outside the SRF cavity bandwidth.  Figure\ref{fig13}  is the cavity probe spectrum for the 30 dB dynamic range.  Figure\ref{fig13} shows the wideband magnitude signal with power spread into the Bessel sidebands. Carrier frequency variation is due to microphonics.  The FM loop around the generator and system follows the resonant frequency, hence the variation in peak marker frequencies of 50-75 Hz.  Other close in sidebands are from microphonic sources.
\section{Conclusions}
A CW SRF cavity control system at 2.45 GHz has been experimentally developed.  Vector control of an injection locked magnetron has been extensively tested and characterized with a SRF cavity as the load.  Amplitude dynamic range of 30 dB, amplitude stability of 0.3\% r.m.s, and phase stability of 0.26 degrees r.m.s. has been demonstrated. Building a system at a specific frequency of interest to the accelerator community will require significant upfront nonrecurring engineering costs for the development of a new magnetron and power supply.  A number of new accelerator applications will require tens to hundreds of such RF power sources.  In these quantities, the expected cost per watt stated in the introduction should be achievable, not to mention electricity savings due to increased efficiency that this RF power source will incur over the life of the accelerator.

\section*{Acknowledgment}
Thanks to Robert Rimmer and Haipeng Wang of JLAB for the loan of the 2.45 GHz single cell SRF cavity, to Elvin Harms and Elias Lopez for setting up the SRF vertical test stand, to Wes Mueller and Pete Seifrid for their assistance with the microwave hardware integration, and finally to Accelerator Division Head Sergei Nagaitsev who supported us from the onset of the effort.

%
%
%
%
\end{document}